\documentclass[a4paper,11pt]{article}
\usepackage{pos}
\usepackage{subfig}

\title{New results on semileptonic $B$ decays from the Belle experiment}
\author*[1]{Lu Cao}
\affiliation{Deutsches Elektronen-Synchrotron (DESY),\\
 Notkestraße 85, Hamburg, Germany}
  
\note{On behalf of the Belle collaboration.}
\emailAdd{lu.cao@desy.de}
\abstract{This proceeding summarizes the recent measurements of semileptonic $b\to u$ decays from the Belle experiment. These analyses use the full data set collected at the $\Upsilon(4S)$ resonance at KEKB accelerator complex with a center-of-mass energy of $\sqrt{s} = 10.58$ GeV, which contains an integrated luminosity of 711 fb$^{-1}$ and corresponds to 772 million $\Upsilon(4S) \to B\bar{B}$ events. In the sector of inclusive decay, the partial branching fraction of $B \to X_{u} \ell \nu$ decays and the CKM matrix element $|V_{ub}|$ are measured with hadronic tagging. The first measurement of differential branching fractions of $B \to X_{u} \ell \nu$ are shown as a function of different kinematic variables in this contribution as a preliminary result. For the exclusive mode, the branching fraction of $B^{+} \to \pi^{+}\pi^{-} \ell \nu$ including both resonant and non-resonant contributions is measured for the first time. The preliminary results on the branching fractions of $B^{+} \to \eta \ell \nu$ and $B^{+} \to \eta^{(\prime)} \ell \nu$ in the full $q^{2}$ range are also presented.}

\FullConference{%
  *** The European Physical Society Conference on High Energy Physics (EPS-HEP2021), ***\\
  *** 26-30 July 2021 ***\\
  *** Online conference, jointly organized by Universität Hamburg and the research center DESY ***
}

\begin{document}

\maketitle

%%%%%%%%%%%%%%%%%%%%%%%
\section{Partial branching fractions of $B \to X_{u} \ell \nu$ and $|V_{ub}|$}
\label{sec:myprd}
%%%%%%%%%%%%%%%%%%%%%%%
One of the crucial tests of the Standard Model of particle physics (SM) is a precise determination of the magnitude of the Cabibbo-Kobayashi-Maskawa (CKM) matrix elements describing the quark mixing and accounts for $CP$-violation in the quark sector. However, the disagreement between the exclusive and inclusive determinations of $|V_{ub}|$ is about three standard deviations \cite{Amhis:2019ckw}. On the other hand, the experimental measurement of the inclusive semileptonic decay $B\to X_{u} \ell \nu$ is challenging due to the large background from the CKM-favoured $B\to X_{c} \ell \nu$ decay. This analysis \cite{cao:2021prd} is motivated to measure the partial branching fractions of three accessible $B \to X_{u} \ell \nu$ phase-space regions and extract the inclusive $|V_{ub}|$.

The signal $B\to X_{u} \ell \nu$ MC sample is a combination of resonances and non-resonant decay using a hybrid modelling approach \cite{hybrid, Prim:2019gtj}. The non-resonant component is based on the theory calculation of Ref.~\cite{DeFazio:1999ptt} with the model parameters in the Kagan-Neubert scheme from Ref.~\cite{Buchmuller:2005zv}. The hadronic decays of one of the $B$ mesons are reconstructed via the full reconstruction algorithm \cite{Feindt:2011mr} based on neural networks. The $B_{\mathrm{tag}}$ reconstruction efficiency is calibrated using a data-driven approach described in Ref.~\cite{Glattauer:2015teq}. All tracks and clusters not used in the construction of the $B_{\mathrm{tag}}$ candidate are used to reconstruct the signal side. The four-momentum of hadronic system $p_{X}$ is defined as a sum of the four-momenta of tracks and clusters which are not involved in reconstructing the $B_{\mathrm{tag}}$ and signal lepton. With the fully reconstructed four-momentum of $B_{\mathrm{tag}}$ and the known beam-momentum, the four-momentum of signal $B$ can be defined. In addition, the signal lepton with $E_{\ell}^{B}=\left|\mathbf{p}_{\ell}^{\mathbf{B}}\right|>1$ GeV in the signal-$B$ rest frame is used to identify the semileptonic decays. The missing mass squared $\mathit{MM}^{2}$ and the four-momentum transfer squared $q^{2}$ are defined as 
\begin{equation}\label{eq:psig}
p_{\mathrm{sig}}=p_{e^{+}e^{-}}-\left(\sqrt{m_{B}^{2}+\left|\mathbf{p}_{\mathrm{tag}}\right|^{2}},\;
\mathbf{p}_{\mathrm{tag}}\right), \;\;
\mathit{MM}^{2}=\left(p_{\mathrm{sig}}-p_{X}-p_{\ell}\right)^{2}, \;\; q^{2}=\left(p_{\mathrm{sig}}-p_{X}\right)^{2}. 
\end{equation}

\begin{table}[b]
\renewcommand\arraystretch{1.0}
\centering
\begin{tabular}{lccc}
\hline\hline

 Fit variable       & Phase-space region                                                                           & $10^{3}\Delta \mathit{BF}$  \\ \hline
$M_{X}$       & $E^{B}_{\ell}>1$ GeV, $M_{X}<1.7$ GeV                                                   &$1.09 \pm 0.05 \pm 0.08$  \\
$q^{2}$       &  $E^{B}_{\ell}>1$ GeV,  $M_{X}<1.7$ GeV, $q^{2}>8$ GeV$^{2}$            & $0.67 \pm 0.07 \pm 0.10$  \\
 $E^{B}_{\ell}$   &  $E^{B}_{\ell}>1$ GeV, $M_{X}<1.7$ GeV                   &$1.11 \pm 0.06 \pm 0.14$  \\ 
 $E^{B}_{\ell}$   &  $E^{B}_{\ell}>1$ GeV                                                         &$1.69 \pm 0.09 \pm 0.26$       \\
 $M_{X}:q^{2}$  &  $E^{B}_{\ell}>1$ GeV                                                                           &$1.59 \pm 0.07 \pm 0.16$    \\ \hline\hline
\end{tabular}
\caption{The measured partial branching fractions for various phase-space regions. The first uncertainty is statistical and the second one is systematics.}
    \label{tab:fit-PS-data}
\end{table}

To separate the signal $B\to X_{u} \ell \nu$ decay from the background events which are dominated by $B\to X_{c} \ell \nu$, a machine learning based classification with boosted decision trees (BDTs) is utilised. After applying the selection of the BDT classifier, a binned likelihood fit is performed to extract the signal yield, where the systematic uncertainties are incorporated via nuisance-parameter constraints. In total, five separate fits are carried out to measure the three partial branching fractions and the results are summarised in Table~\ref{tab:fit-PS-data}. 

The $|V_{ub}|$ is extracted based on the measured partial branching fractions value with the average $B$ meson lifetime of $1.579\pm0.004\,$ps \cite{pdg:2020} and the state-of-the-art theory predictions on decay rate: BLNP \cite{BLNP}, DGE \cite{DGE1,DGE2}, GGOU \cite{GGOU} and ADFR \cite{ADFR1,ADFR2}. The arithmetic average of the most precise determinations for the phase-space region $E^{B}_{\ell}>1$ GeV is $\left|V_{u b}\right|=(4.10 \pm 0.09_{\texttt{stat}} \pm 0.22_{\texttt{syst}} \pm 0.15_{\texttt{theo}}) \times 10^{-3}$. The compatibility with the world averages of exclusive results $\left|V_{u b}^{\text {excl.}}\right|=(3.67 \pm 0.09 \pm 0.12) \times 10^{-3}$ \cite{Amhis:2019ckw} is 1.3 standard deviations; it is also compatible with the value expected from CKM unitarity from a global fit of Ref.~\cite{Charles:2004jd} of $\left|V_{u b}\right|=(3.62^{+0.11}_{-0.08}) \times 10^{-3}$ within 1.6 standard deviations.

\begin{figure}[t]
	\centering
	    \includegraphics[width=0.5\linewidth]{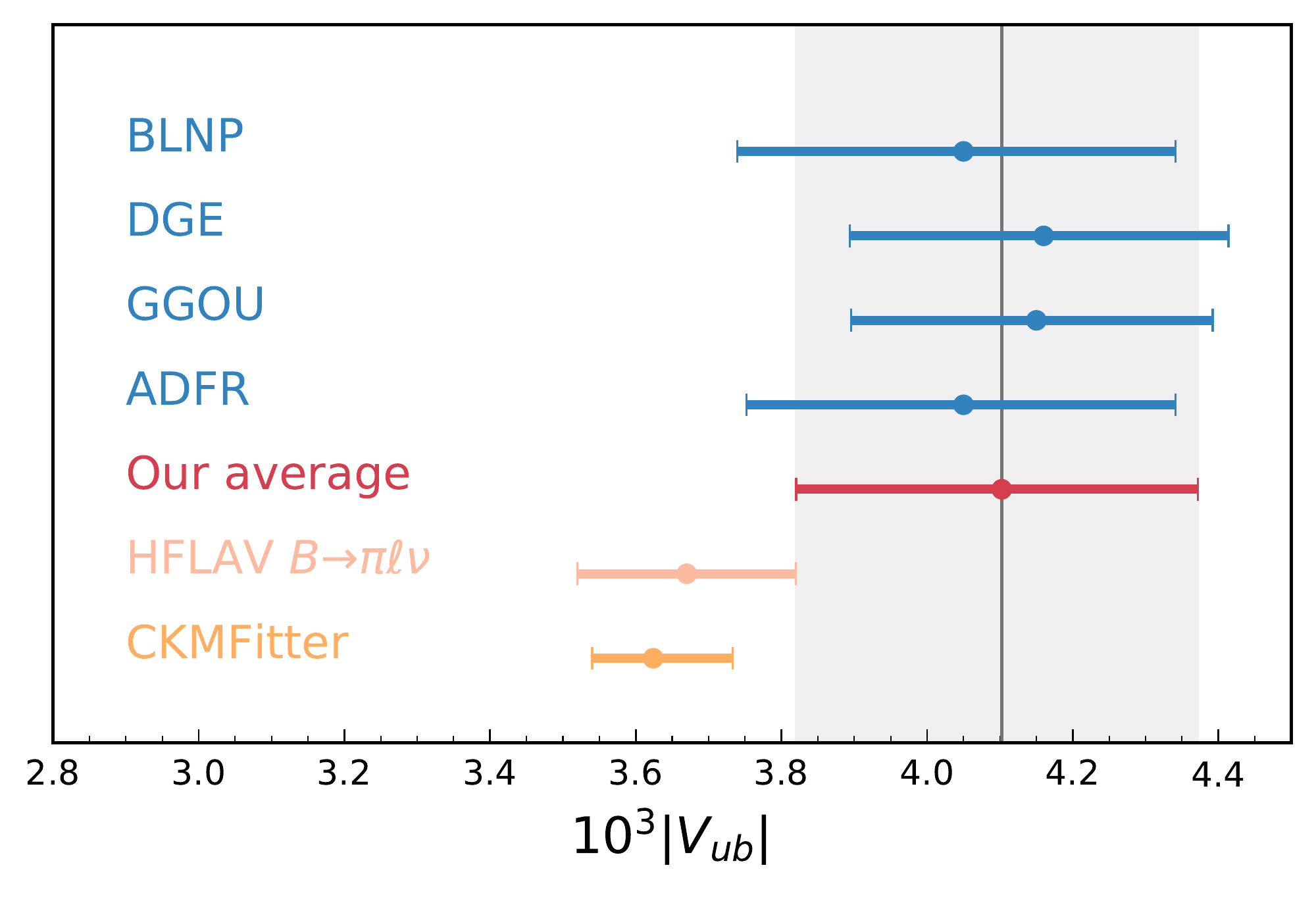}
	\caption{\label{fig:vub} The extracted inclusive $|V_{ub}|$ based on the four calculations
and the arithmetic average is compared to the determination from exclusive method \cite{Amhis:2019ckw} and the expectation from CKM unitarity \cite{Charles:2004jd}.}
\end{figure}

%%%%%%%%%%%%%%%%%%%%%%%
\section{Differential branching fractions of $B \to X_{u} \ell \nu$}
\label{sec:myprl}
%%%%%%%%%%%%%%%%%%%%%%%
The first measurements of differential spectra of inclusive $B \to X_{u} \ell \nu$ decays \cite{cao:2021prl} are reported as a function of the lepton energy $E^{B}_{\ell}$ in the $B$ rest frame, the four-momentum-transfer squared $q^{2}$, light-cone momenta $P_\pm = \left( E^{B}_X \mp |\mathbf{p}^{B}_X|  \right)$, the hadronic mass $M_{X}$, and the hadronic mass squared $M_{X}^{2}$. The event reconstruction strategy of this analysis is the same as applied for the partial branching fraction measurement in Ref.\cite{cao:2021prd}. To improve the signal purity and
reduce background shape uncertainty, additional selections with $\big|E_{\mathrm{miss}} - |\mathbf{p}_{\mathrm{miss}}|\big| < 0.1 \, \mathrm{GeV}$ and the reconstructed $M_{X} < 2.4 \, \mathrm{GeV}$ are included. The remaining backgrounds are subtracted by fitting the $M_{X}$ distribution after all selections. The detector effects on resolution and acceptance are corrected by unfolding the background subtracted spectra using a singular-value-decomposition (SVD) method \cite{Hocker:1995kb}. The full background subtraction uncertainties and correlations are propagated through the unfolding procedure. The unfolded signal is further converted to differential branching fractions and corrected for reconstruction efficiency and phase-space acceptance. 

\begin{figure}[t]
	\centering
	    \includegraphics[width=0.32\linewidth]{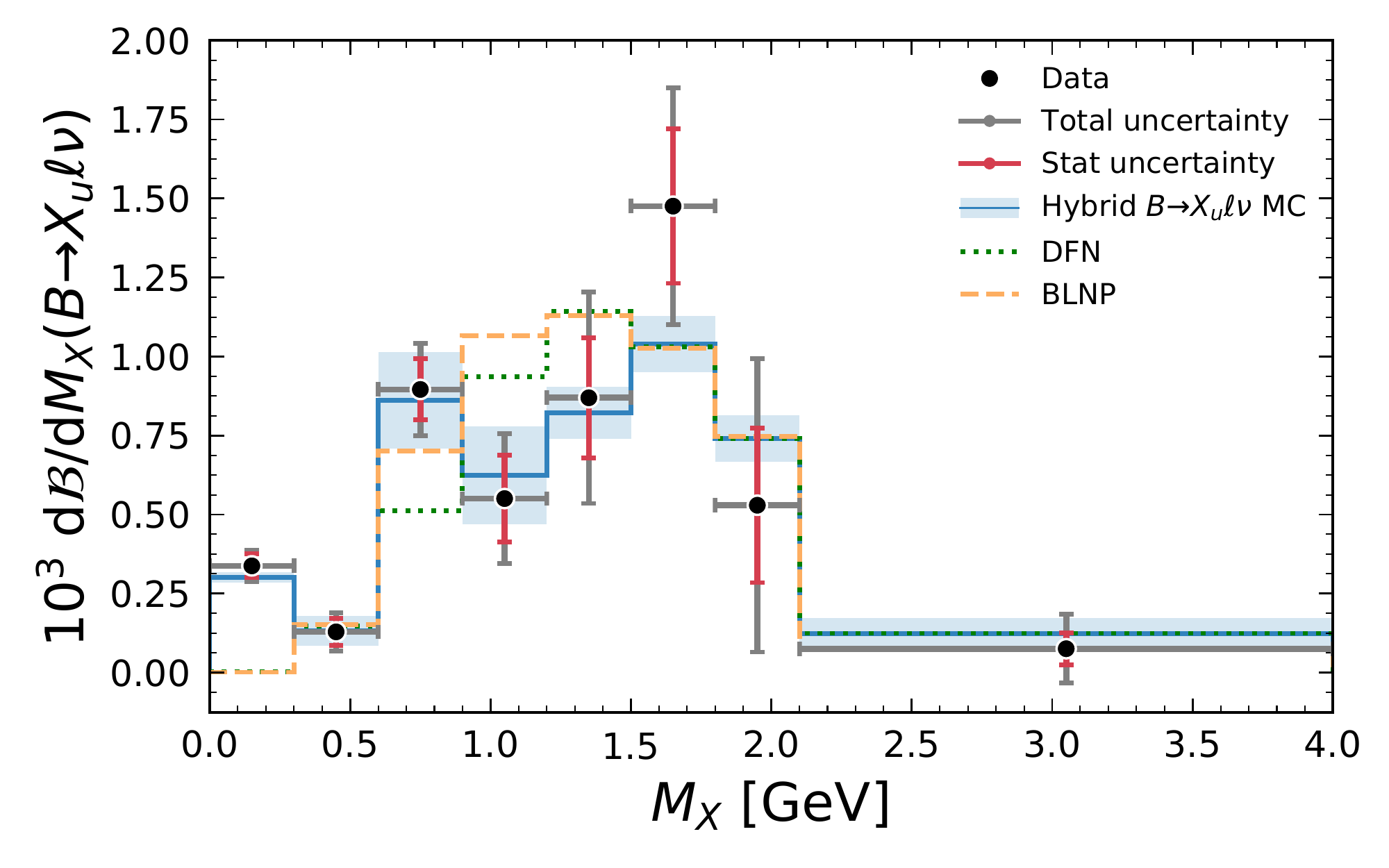}
	    \includegraphics[width=0.32\linewidth]{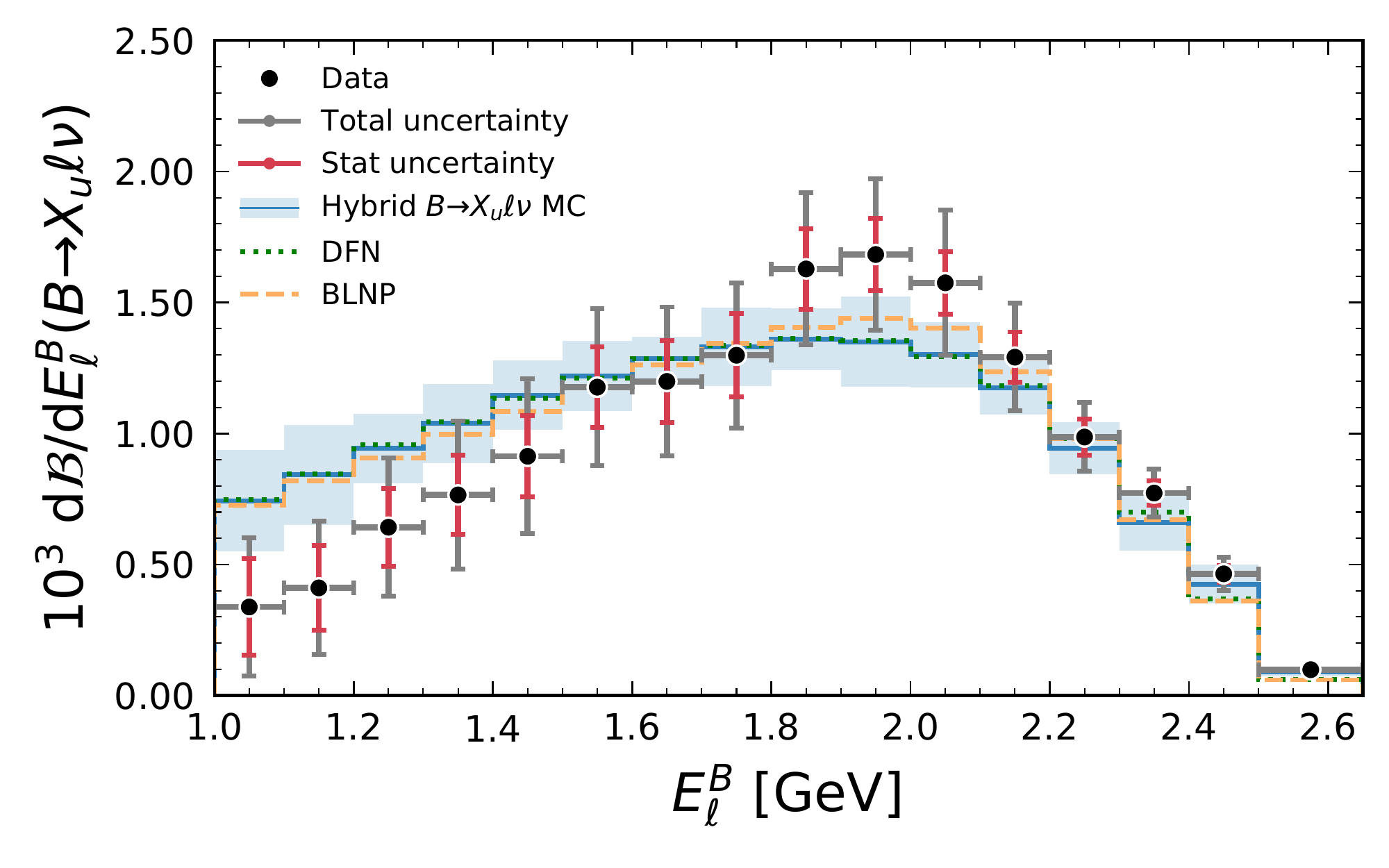}
	    \includegraphics[width=0.32\linewidth]{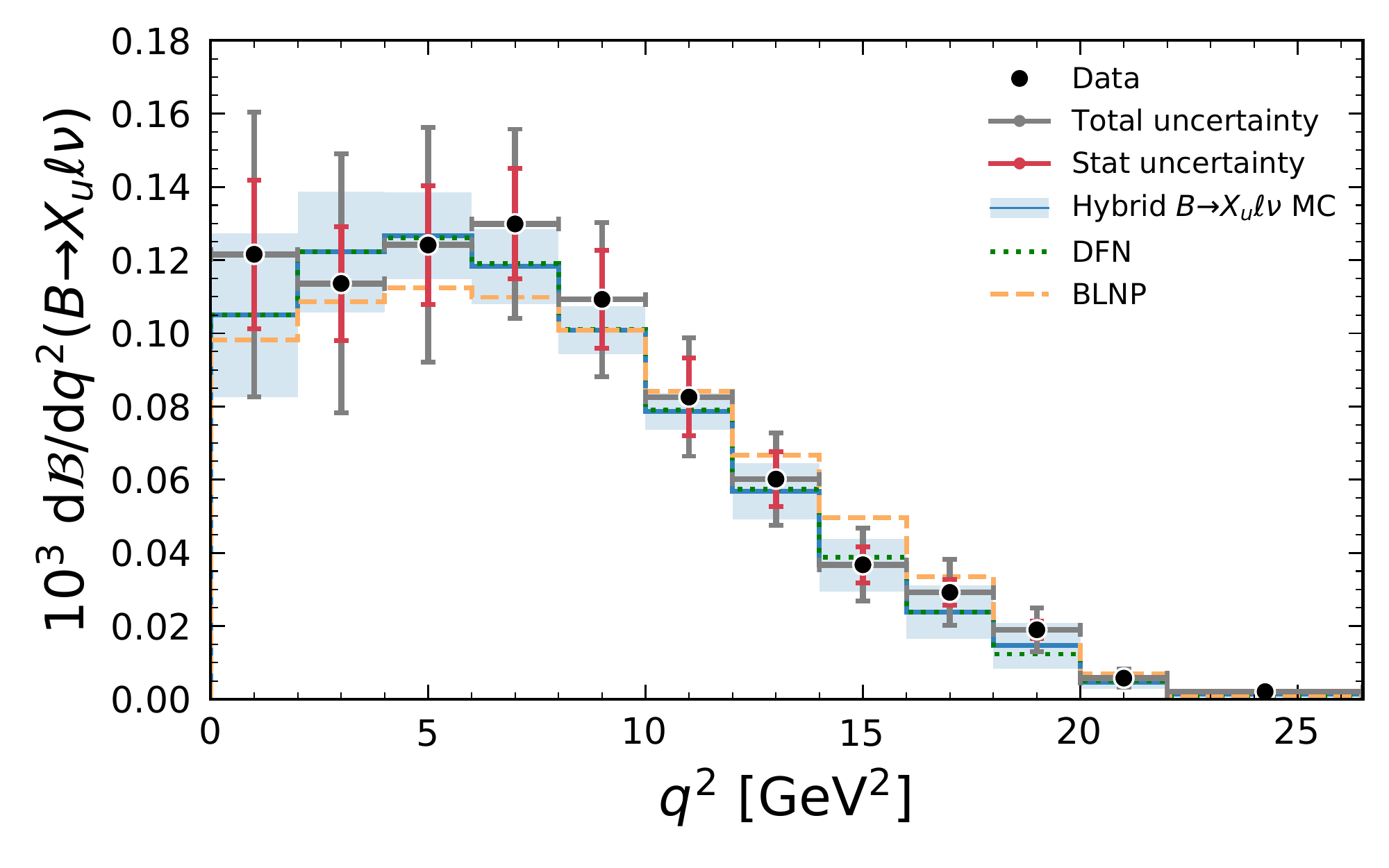}
	    \includegraphics[width=0.32\linewidth]{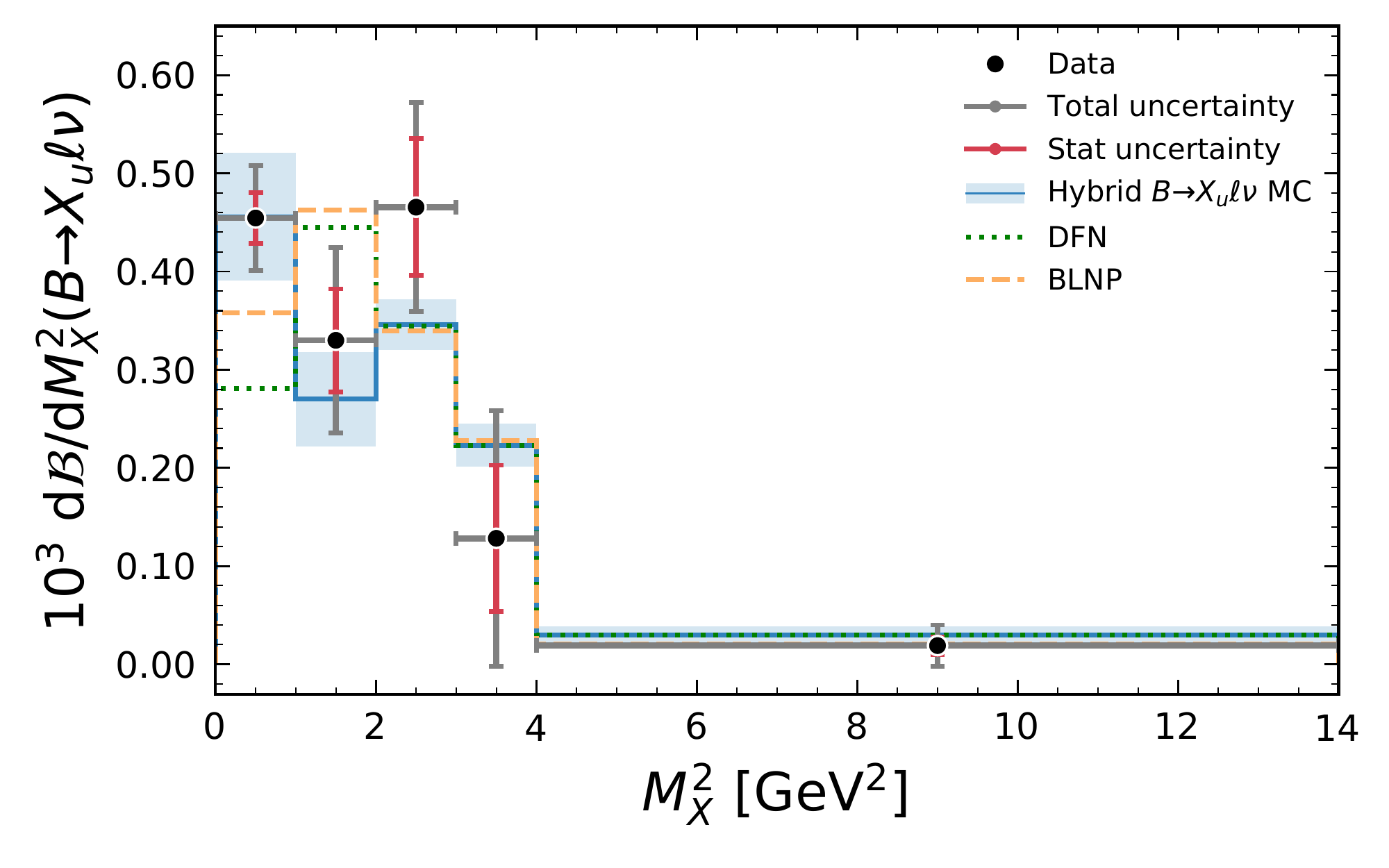}
	    \includegraphics[width=0.32\linewidth]{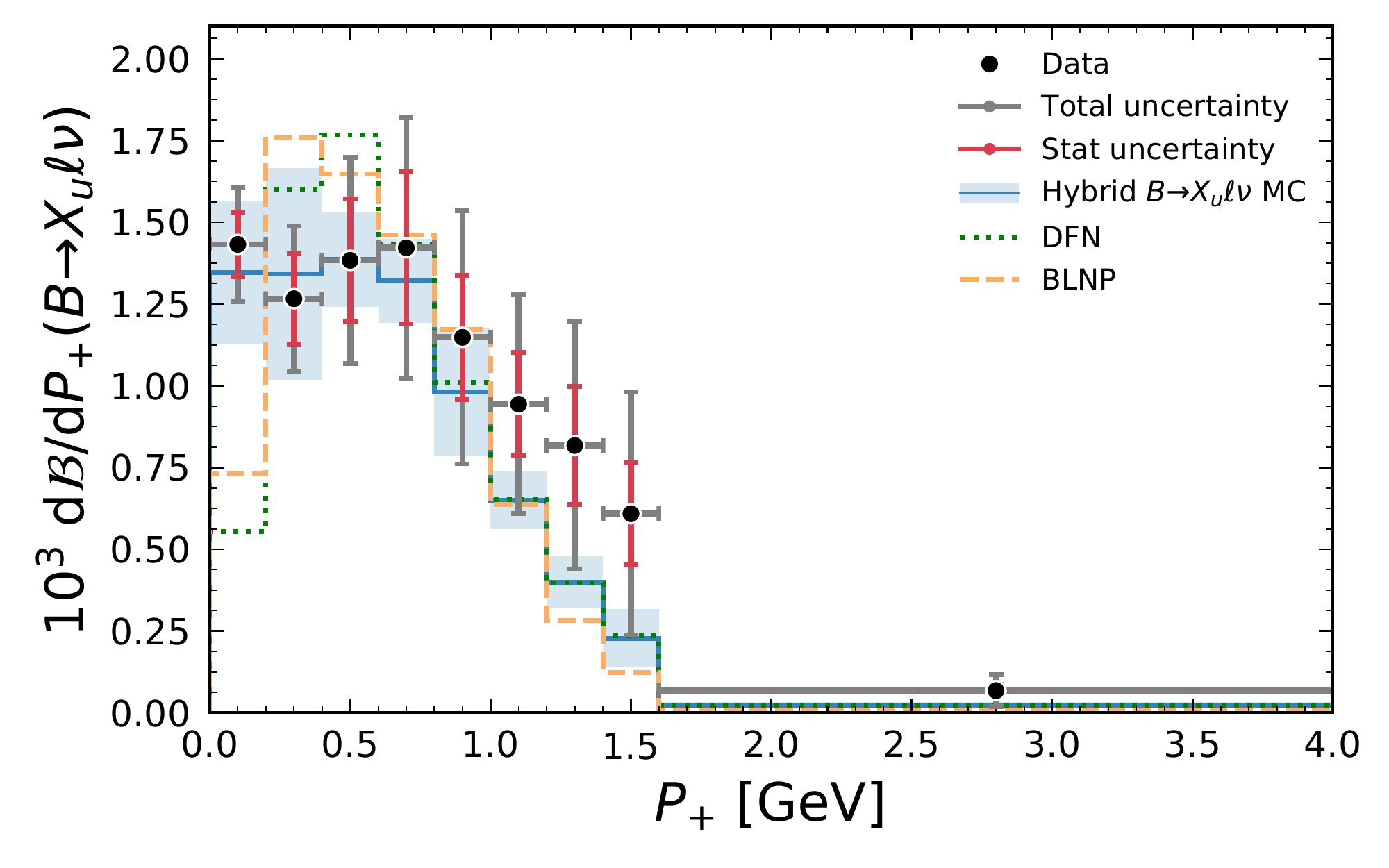}
	    \includegraphics[width=0.32\linewidth]{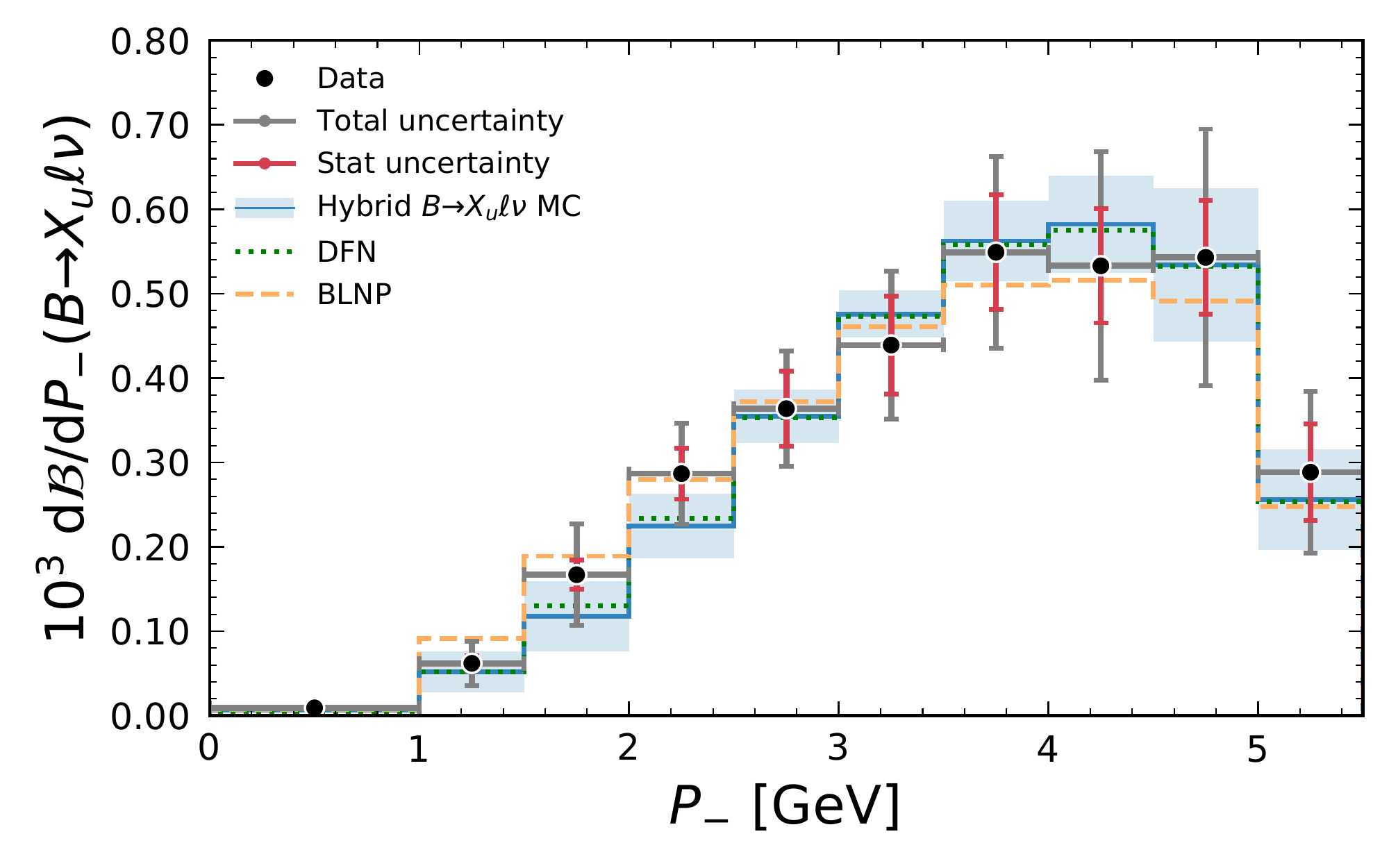}
	\caption{\label{fig:dBF} The measured differential $B \to X_{u} \ell \nu$ branching fractions. The hybrid MC prediction and two inclusive calculations are also shown and scaled to $\Delta \mathcal{B}= 1.59 \times 10^{-3}$.}
\end{figure}

The measured differential $B \to X_{u} \ell \nu$ branching fractions are shown in Fig.~\ref{fig:dBF}.
The measurements show a fair agreement with hybrid and inclusive predictions in general. The hybrid MC describes the $B \to X_{u} \ell \nu$ process more adequately due to the explicit inclusion of resonant contributions. The numerical values with full correlations of measured differential spectra are also provided in Ref.\cite{cao:2021prl}. This result paves the way for future direct determinations of the shape function and $|V_{ub}|$, as proposed by Refs.~\cite{Bernlochner:2020jlt,Gambino:2016fdy}. These novel analyses will provide new insights into the persistent tensions on the value of $|V_{ub}|$ from inclusive and exclusive determinations.

%%%%%%%%%%%%%%%%%%%%%%%
\section{Branching fraction of $B^{+} \to \pi^{+}\pi^{-} \ell^{+} \nu$}
\label{sec:pipi}
\begin{figure}[b]
	\centering
	    \includegraphics[width=0.32\linewidth]{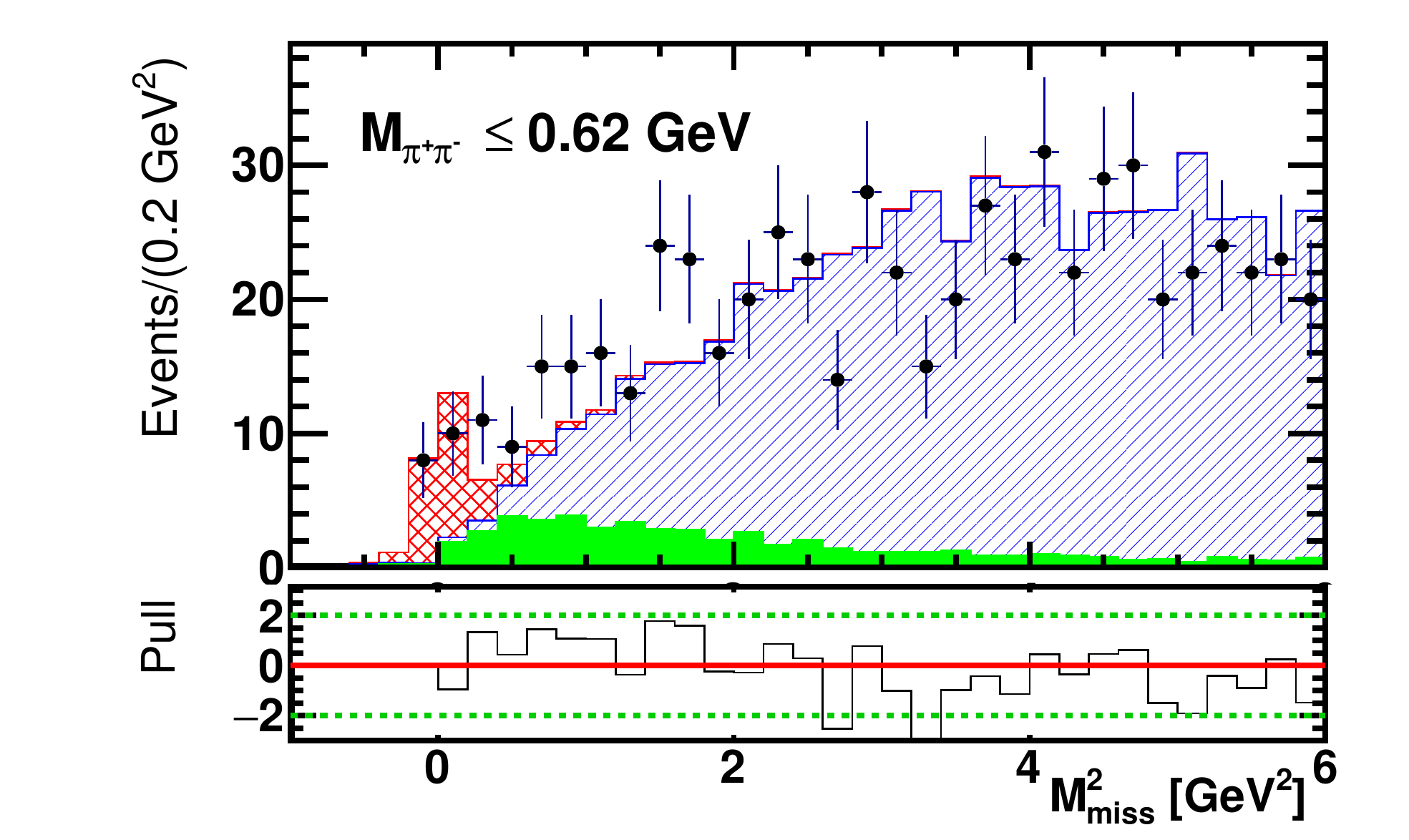}
	    \includegraphics[width=0.32\linewidth]{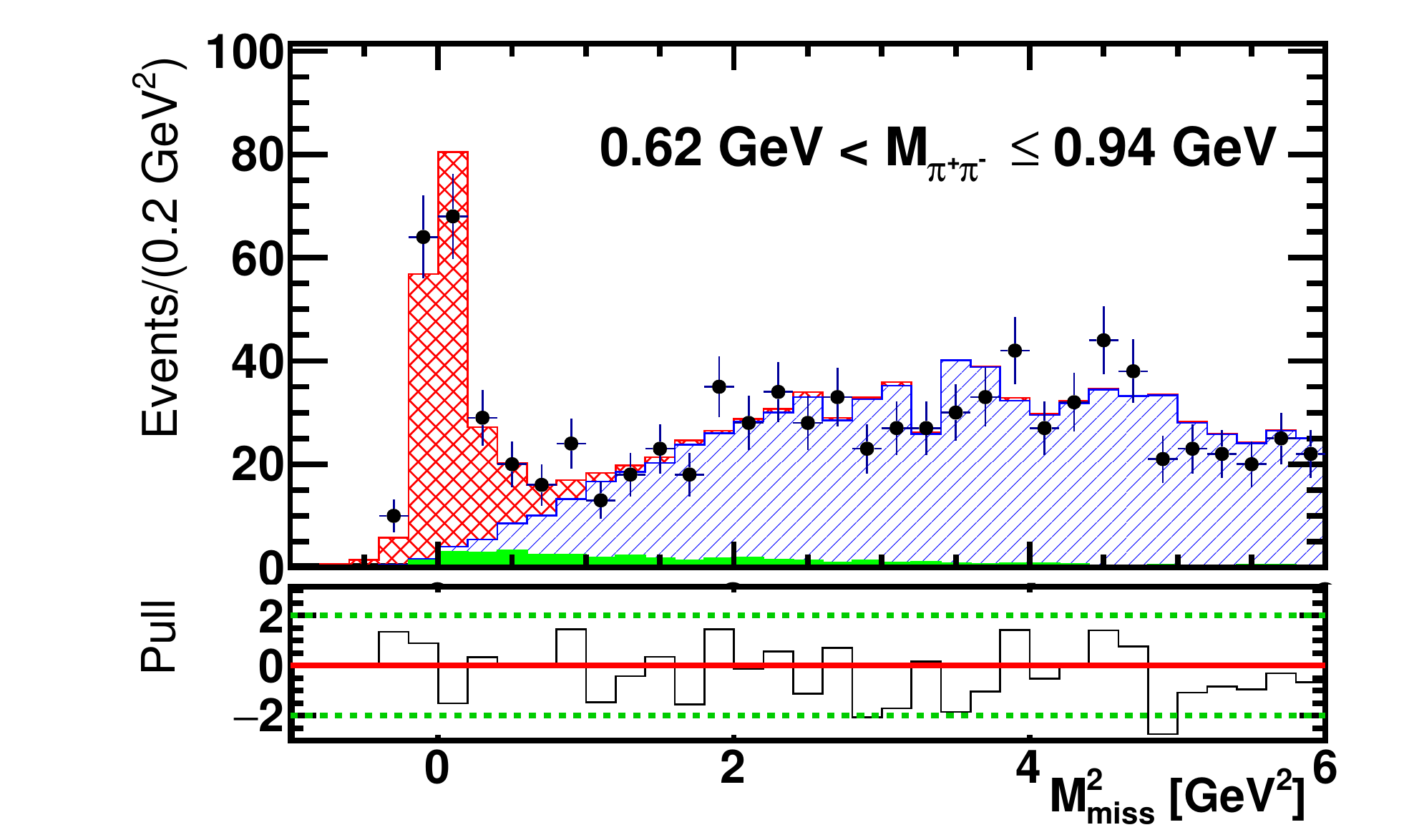}
	    \includegraphics[width=0.32\linewidth]{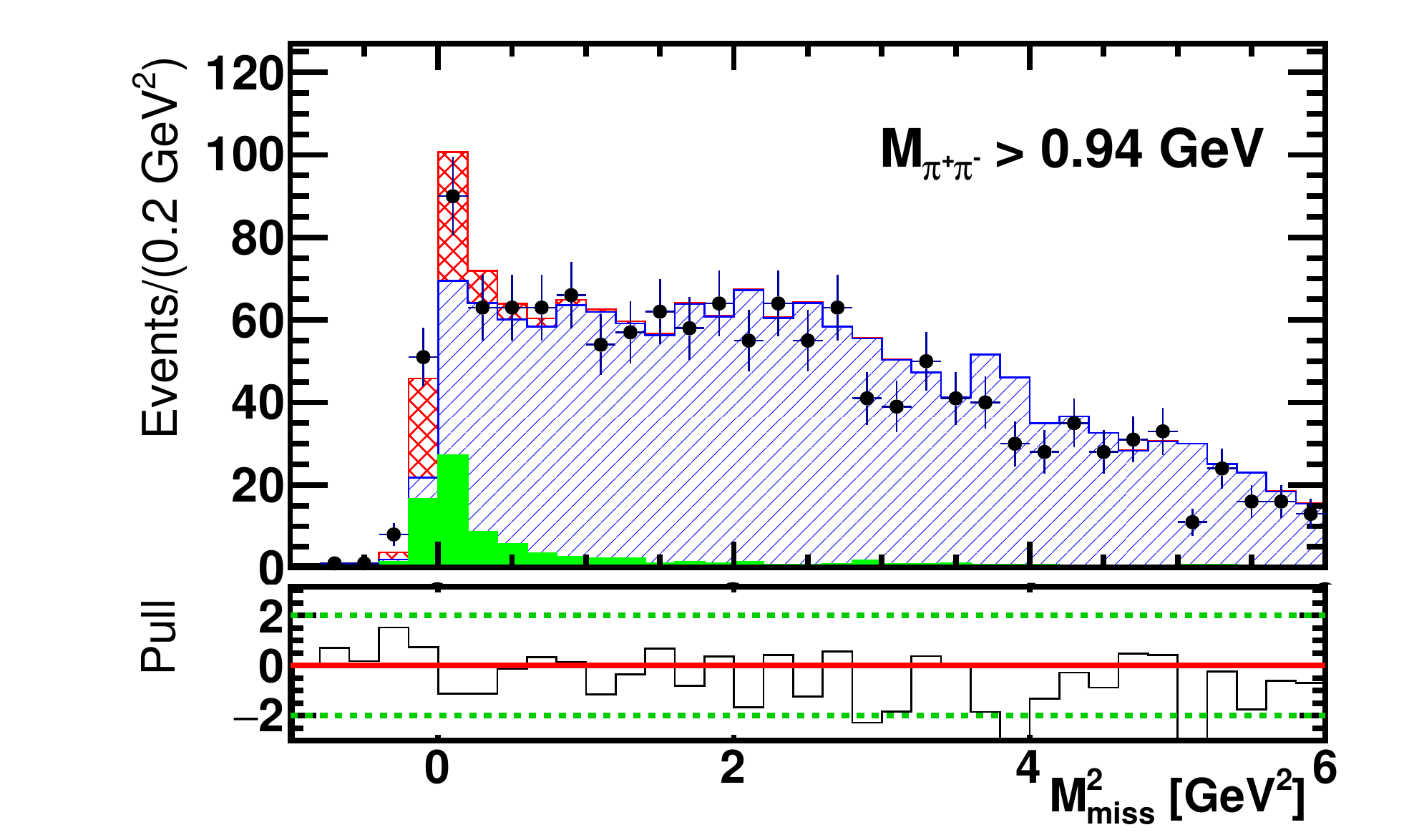}
	\caption{\label{fig:pipi} The projection of the 1D-fit of $M_{\pi\pi}$ result in the $M_{\mathrm{miss}}^{2}$ distribution (points with error bars) in three regions of the dipion mass. The fit components include the signal $B^{+} \to \pi^{+}\pi^{-} \ell^{+} \nu$ (red), $B \to X_{c} \ell^{+} \nu$ background (blue) and fixed background (green).}
\end{figure}

%%%%%%%%%%%%%%%%%%%%%%%
The exclusive $B^{+} \to \pi^{+}\pi^{-} \ell^{+} \nu$ decay is measured with hadronic tagging technique and reported in Ref.\cite{Belle:pipi}. This analysis utilizes the charged $B_{\mathrm{tag}}$ to reconstruct the missing momentum as $P_{\mathrm {miss}}=P_{\Upsilon(4 S)}-P_{B_{\mathrm{tag}}^{\pm}}-P_{\ell^{\mp}}-P_{\pi^{+}}-P_{\pi^{-}}$. The background suppression is based on a BDT trained on six kinematic variables which are effective against other $B$ decays and continuums. After selecting on the BDT classifier output, a binned extended maximum-likelihood fit is performed to the missing mass squared $M_{\mathrm{miss}}^{2}$ spectrum to extract signal. The signal yields are measured in bins of $M_{\pi\pi}=\sqrt{\left(P_{\pi^{+}}+P_{\pi^{-}}\right)^{2}}$ and $q^{2}=\left(P_{\ell}+P_{\nu_{\ell}}\right)^{2}$ using three fit configurations to allow for a $B^{+} \rightarrow \pi^{+} \pi^{-} \ell^{+} \nu_{\ell}$ decay-model-independent interpretation of the result. The three configurations include the 1D-fit respectively on $M_{\pi\pi}$ and $q^{2}$, and a 2D-fit combines these. The projection of the 1D-fit of $M_{\pi\pi}$ result in the $M_{\mathrm{miss}}^{2}$ distribution is shown in Fig.~\ref{fig:pipi}.

With each fit configuration, the extracted signal yields are converted to the total branching fraction by summing over all bins of partial branching fractions and corrected for the corresponding reconstruction efficiency and acceptance. The experimental uncertainty is dominated
by systematic uncertainties in all three configurations. The result from the 1D-fit of $q^{2}$ lying in the middle of the three configurations and is taken as the final result: $\mathcal{B}\left(B^{+} \rightarrow \pi^{+} \pi^{-} \ell^{+} \nu_{\ell}\right)=\left[22.7_{-1.6}^{+1.9}(\text { stat }) \pm 3.5(\text { syst })\right] \times 10^{-5}$. This analysis is the first measurement of the full $M_{\pi \pi}$ spectrum in $B^{+} \to \pi^{+}\pi^{-} \ell^{+} \nu$ decay including both of the resonant and non-resonant contributions.

%%%%%%%%%%%%%%%%%%%%%%%
\section{Branching fractions of $B^{+} \to \eta \ell^{+} \nu$ and $B^{+} \to \eta^{\prime} \ell^{+} \nu$}
\label{sec:eta}
\begin{figure*}[t]
\centering
\subfloat[$M_{\mathrm{bc}} (\eta\to\gamma\gamma)$]{
\includegraphics[width=0.25\textwidth ]{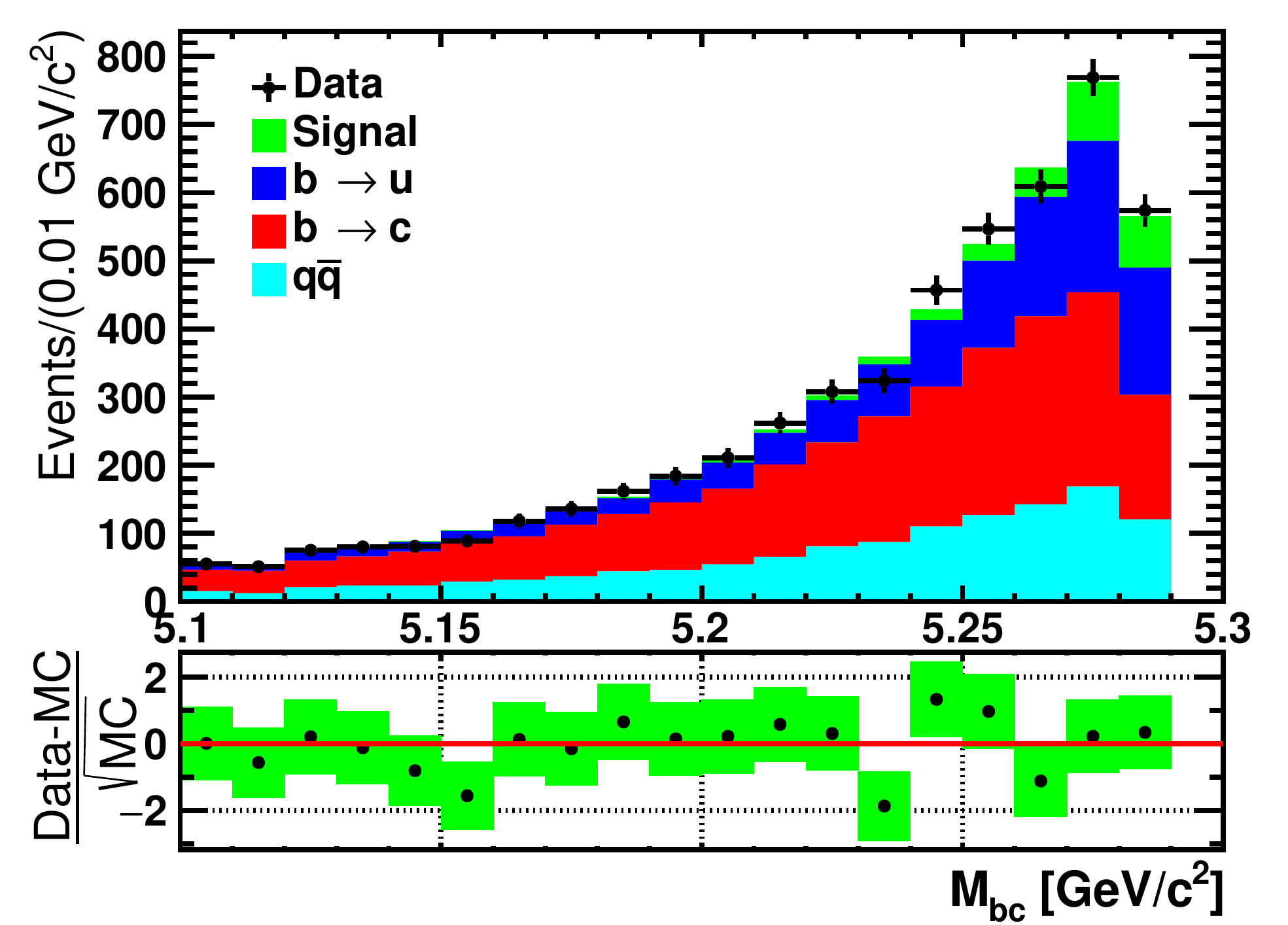}
}
\subfloat[$M_{\mathrm{bc}} (\eta\to\pi^+\pi^-\pi^0)$]{
\includegraphics[width=0.25\textwidth ]{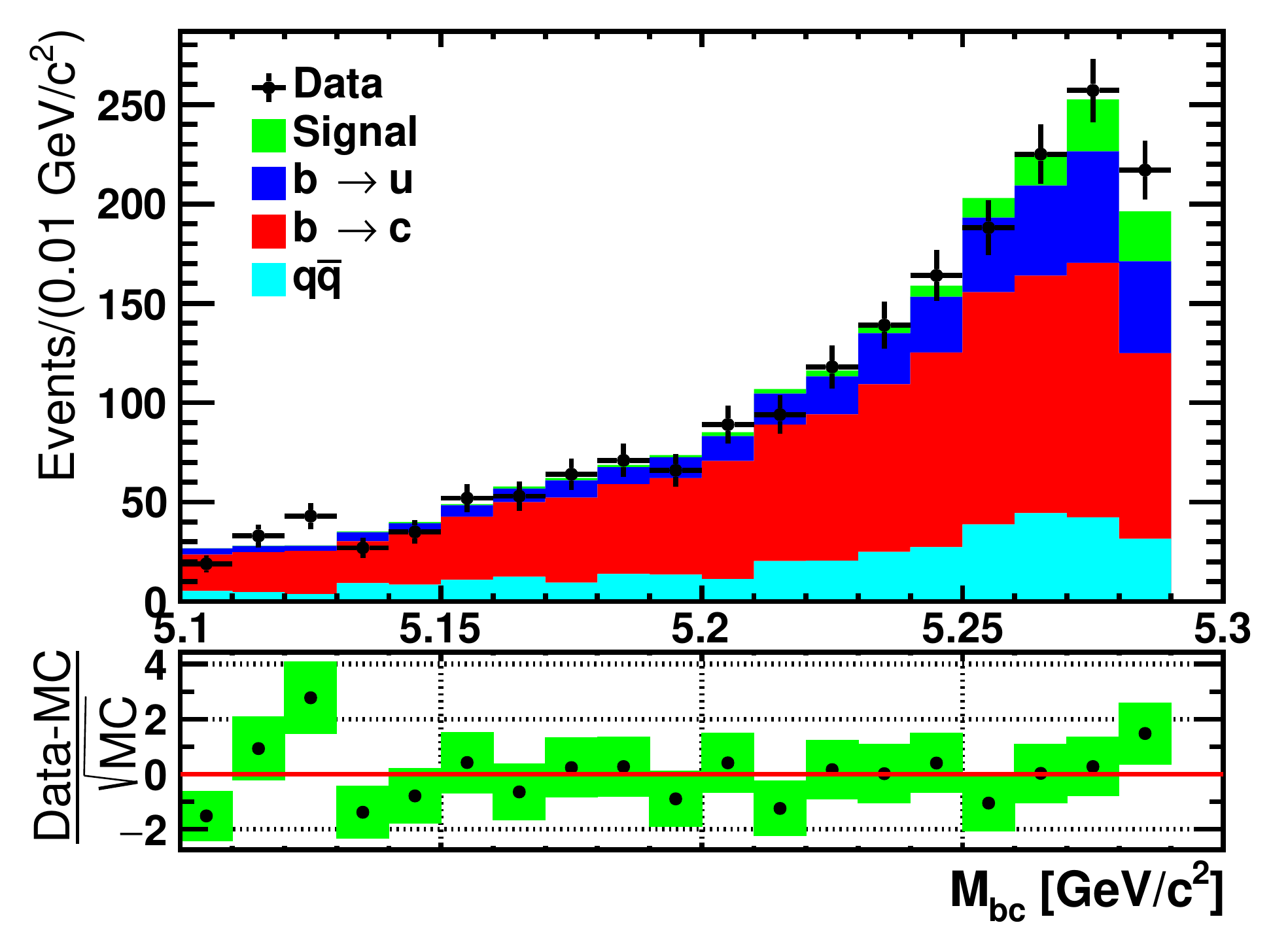}
}
\subfloat[$M_{\mathrm{bc}} (\eta^\prime\to\pi^+\pi^-\eta(\gamma\gamma))$]{
\includegraphics[width=0.25\textwidth ]{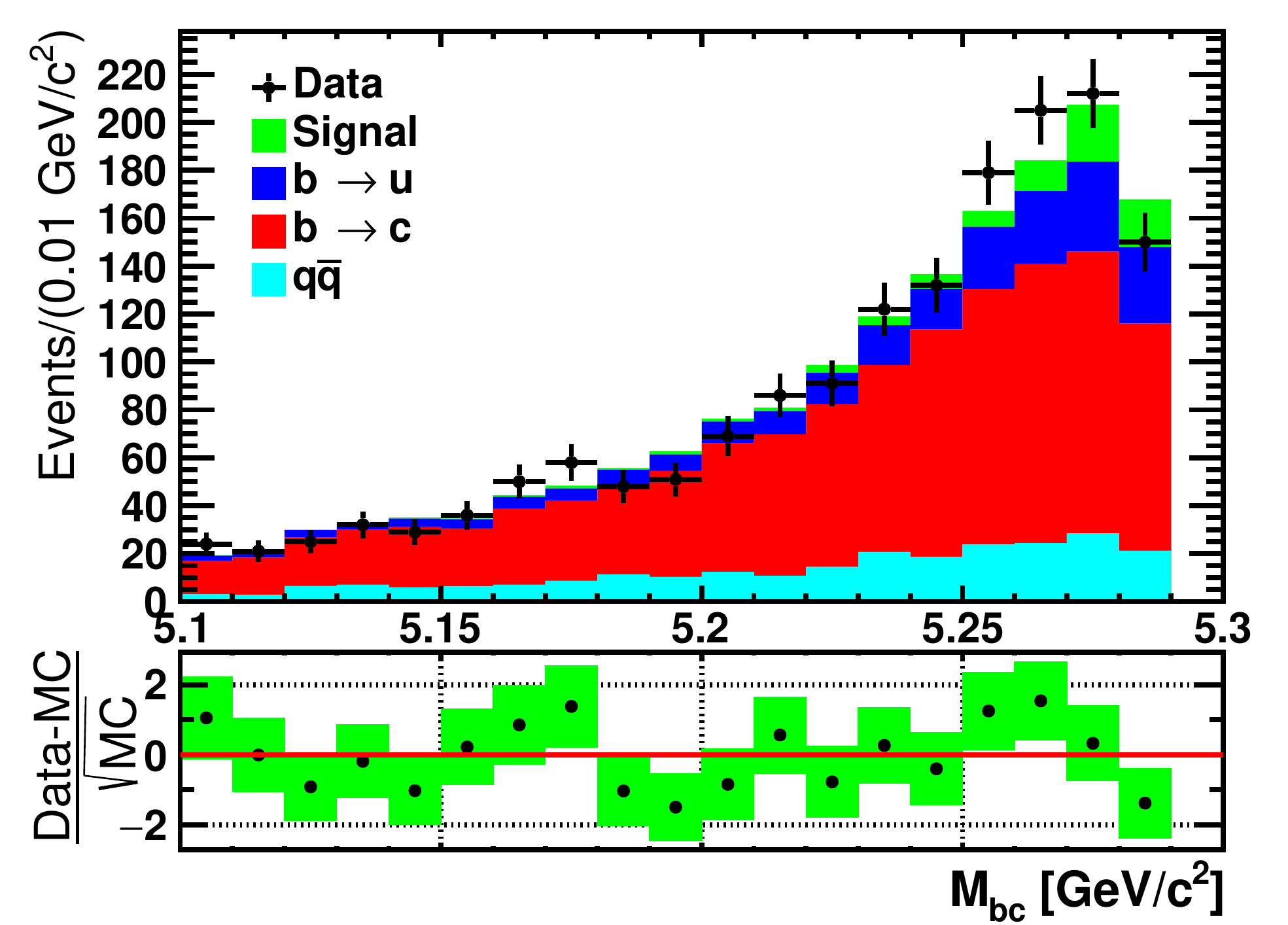}
}\\
\subfloat[$\Delta E (\eta\to\gamma\gamma)$]{
\includegraphics[width=0.25\textwidth ]{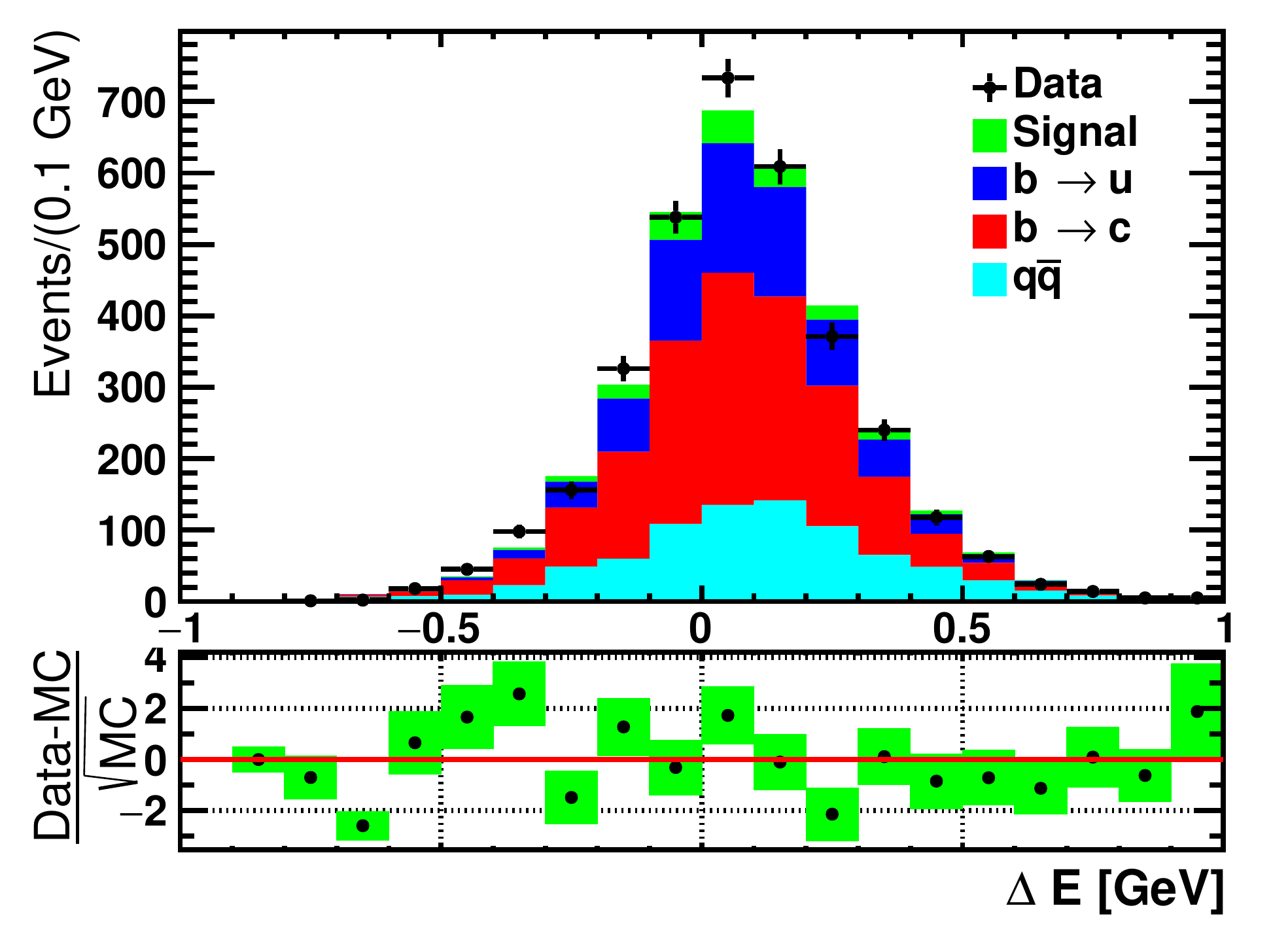}
}
\subfloat[$\Delta E (\eta\to\pi^+\pi^-\pi^0)$]{
\includegraphics[width=0.25\textwidth ]{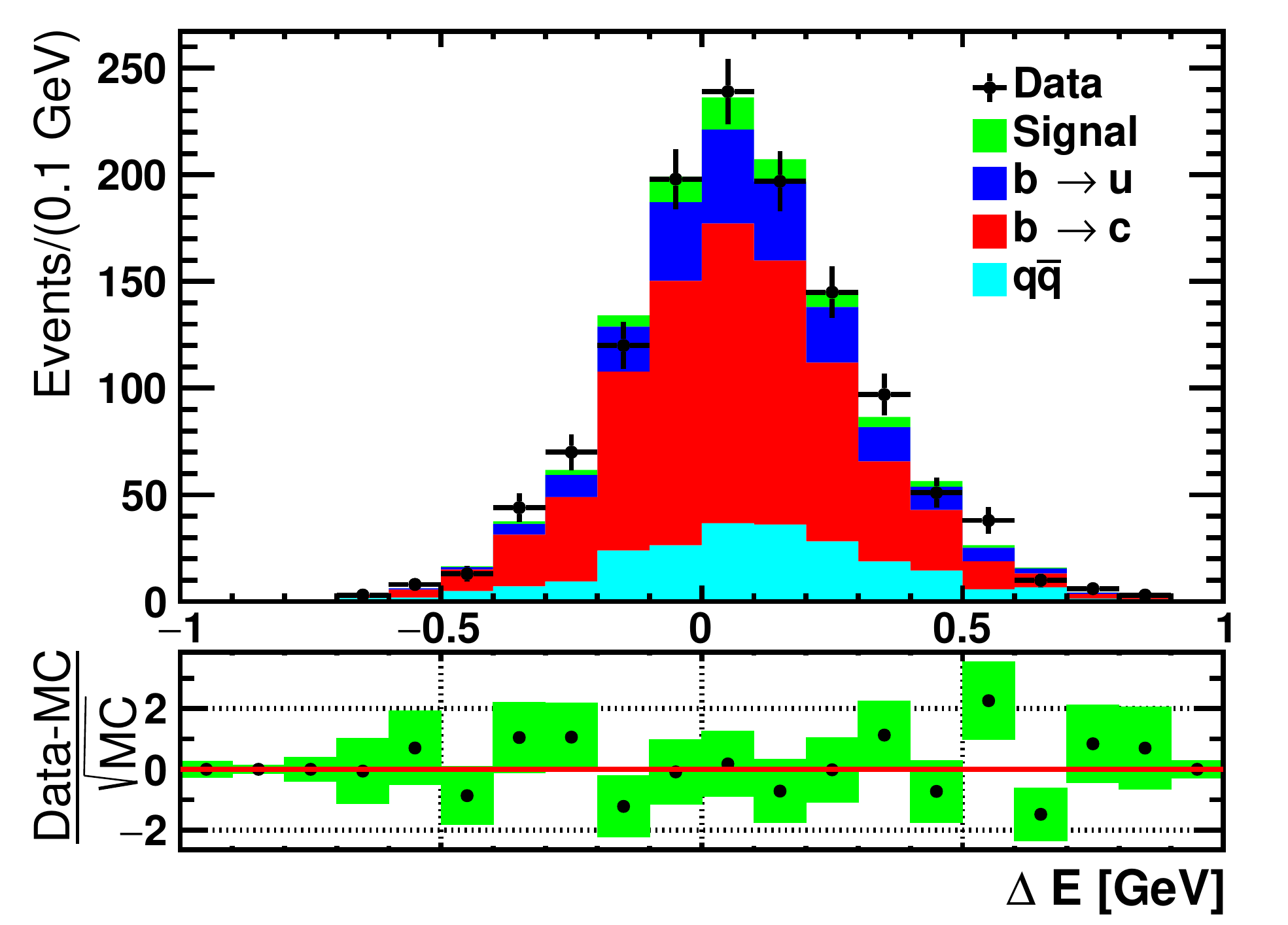}
}
\subfloat[$\Delta E (\eta^\prime\to\pi^+\pi^-\eta(\gamma\gamma))$]{
\includegraphics[width=0.25\textwidth ]{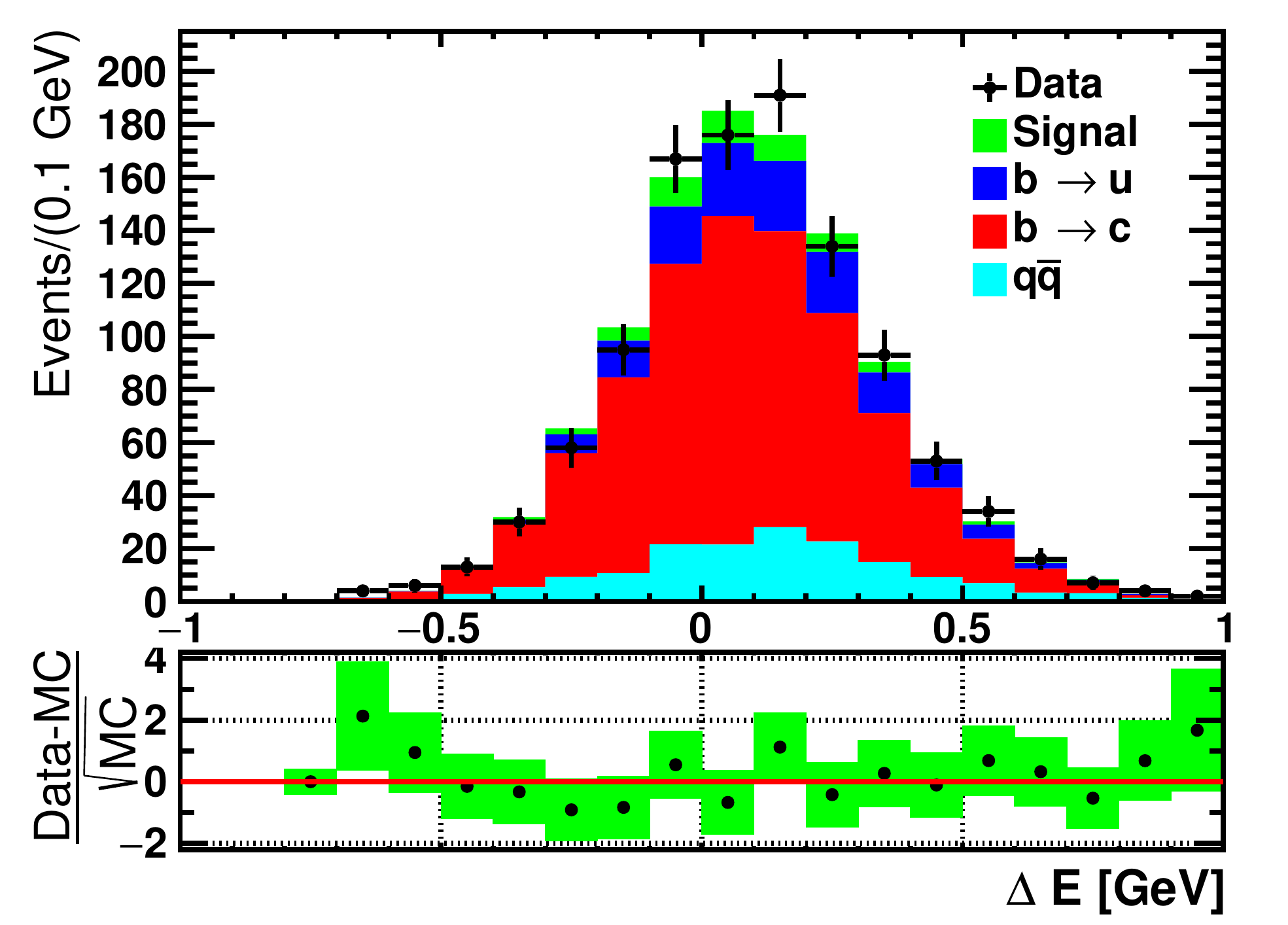}
}
\caption{Projections onto the two fit variables for all three channels used, with the contributions scaled to those obtained in the fit. The other variable is restricted to a signal-enhanced region for visibility.}\label{fig:eta}
\label{fitprojections}
\end{figure*}

%%%%%%%%%%%%%%%%%%%%%%%
The branching fractions of $B^{+} \to \eta^{(\prime)} \ell^{+} \nu$ are measured in the full $q^{2}$ range with untagged method and the preliminary result is reported in Ref.~\cite{Belle:eta}. The decay modes of $\eta$ include $\eta \to \gamma\gamma$ and $\eta \to \pi^{+}\pi^{-}\pi^{0}$. For $\eta^{\prime}$, the di-photon mode of $\eta$ is combined with a pair of pions, i.e. $\eta^{\prime} \to \pi^{+}\pi^{-}\eta$. The invariant mass and a combined mass-vertex fit are useful to suppress backgrounds. Further background subtraction is based on the angle between $B$ meson and the visible final state $|\cos (\theta_{B \ell \eta^{(\prime)}}^{*})|<1$ to ensure the signal events reconstructed within the physical region. The events with missing mass squared $|m^{2}_{\mathrm{miss}}| > 7 \, \text{GeV}^{2}$ are further rejected. After all selections, the signal yields are extracted by a binned maximum-likelihood fit using the beam-constrained mass $M_{\mathrm{bc}}=\sqrt{E_{\mathrm{beam}}^{* 2} -\vec{p}_{B}^{* 2}}$ and the energy difference $\Delta E=E_{B}^{*} - E_{\mathrm{beam}}^{*}$. The projections of fit results are shown in Fig.~\ref{fig:eta}. The resulting branching fractions are $\mathcal{B}\left(B^{+} \rightarrow \eta \ell^{+} \nu_{\ell}\right)=$ $\left(2.83 \pm 0.55_{\text {(stat.) }} \pm 0.34_{\text {(syst. })}\right) \times 10^{-5}$ and $\mathcal{B}\left(B^{+} \rightarrow \eta^{\prime} \ell^{+} \nu_{\ell}\right)=\left(2.79 \pm 1.29_{\text {(stat.) }} \pm 0.30_{\text {(syst.})}\right) \times 10^{-5}$.

%%%%%%%%%%%%%%%%%%%%%%%
\section{Summary}
%%%%%%%%%%%%%%%%%%%%%%%

Several semileptonic $b\to u$ decays based on the full Belle data set are measured recently, which provide fruitful investigations on the B-flavor physics. Beyond these important results, the accumulated knowledge on such as MC modeling improvements, advanced analysis techniques will be beneficial for future measurements e.g. Belle II or LHCb.

\bibliographystyle{apsrev4-1}
\bibliography{ref}

\end{document}